\documentclass[onecolumn]{aastex701}
\usepackage{amsmath}
\usepackage{subcaption}
\usepackage{graphicx}
\usepackage{CJK}
\usepackage{color}

\newcommand{\response}[1]{#1}
\newcommand{\typo}[1]{#1}
\newcommand{\refinement}[1]{#1}
\newcommand{\revisionii}[1]{#1}
\begin{document}
\begin{CJK*}{UTF8}{gbsn} 

\title{\refinement{A Fast Concentric-disk Contour Integration Method For Microlensing Limb-darkening Effect}}

\author[orcid=0009-0008-3668-9045]{Suwei Wang (王苏为)}
\affil{The Kavli Institute for Astronomy and Astrophysics,
  Peking University, Beijing 100871, China}
\affil{Department of Astronomy, School of Physics, Peking
  University, Beijing 100871, China}
\email{suwei\_wang@stu.pku.edu.cn}

\begin{abstract}
Incorporating limb darkening is a computationally demanding step in contour-integration-based microlensing modeling. Conventional concentric-ring integration
\revisionii{is incompatible with high-order quadrature schemes,} limiting efficiency.
\refinement{We develop a new concentric-disk method, which reformulates the limb-darkening integral and enables the application of high-order adaptive quadrature, significantly accelerating the computation.}
\refinement{While mainly demonstrated using a linear limb-darkening profile, the method readily extends to more general limb-darkening profiles.}
As a source effect, it applies to lens systems of any complexity.
The method achieves a convergence rate scaling faster than $N_{\rm{uni}}^{-4}$ with the number of uniform-source magnification evaluations $N_{\rm{uni}}$, a significant improvement over the concentric-ring $N_{\rm{uni}}^{-2}$ scaling. For a relative accuracy of $10^{-6}$, the concentric-disk approach typically requires only $30\%$ or less of the computational cost of traditional algorithms. This method has been implemented in the binary-lens contour integration code \texttt{Twinkle}, providing an efficient and precise tool for analyzing current and future high-precision microlensing observations.

\end{abstract}


\section{Introduction}

With the rapidly increasing rate of microlensing planet detections, computational efficiency and numerical precision in light-curve modeling have become critical for extracting scientific results from current and future surveys \citep{2012ARA&A..50..411G, 2021ARA&A..59..291Z}. Gravitational microlensing remains uniquely sensitive to low-mass exoplanets near and beyond the snow line, with ongoing projects such as the Optical Gravitational Lensing Experiment IV \citep{Udalski2015}, the Korea Microlensing Telescope Network \citep{Kim2016}, and the Microlensing Observations in Astrophysics survey \citep{MOA} having yielded hundreds of planetary detections. The forthcoming Nancy Grace Roman Space Telescope is expected to expand this yield dramatically, particularly for Earth-mass planets whose signals demand exceptionally precise magnification calculations \citep{Penny2019, Yee2023}.

The calculation of finite-source magnification in microlensing primarily relies on two algorithmic approaches: ray-shooting methods \citep{1986A&A...166...36K, SchneiderWeiss1986, 1997MNRAS.284..172W} and contour integration methods \citep{1987A&A...174..361S, 1995A&AS..109..597D}. The ray-shooting approach traces light rays backward from the image plane \revisionii{(the angular coordinate plane of lensed images)} to the source plane \revisionii{(the unlensed angular position plane of the source)}, offering a computationally straightforward method well-suited for modeling complex lens geometries, \revisionii{requiring dense two-dimensional sampling of the image plane.}
In contrast, \revisionii{the contour integration method obtains the magnification from image boundaries using contour integrals.
Depending on the implementation, image contours may be constructed either through solving the lens equation or directly through image-plane sampling.}
This approach has been implemented in several codes, such as 
\texttt{AdaptiveContouring} \citep{AC_2007MNRAS.377.1679D},
\texttt{VBBinaryLensing} \citep{VBBL2010}, 
\texttt{triplelens} \citep{Kuang_triple},
\texttt{VBMicroLensing} \citep{VBML2024}, 
\texttt{microlux} \citep{haibin_microlux}, 
and our recently developed \texttt{Twinkle} code \citep{Twinkle2025}, offering superior flexibility, especially for modeling effects like lens orbital motion.

While the uniform-source approximation provides a useful baseline for many microlensing calculations, real stellar atmospheres are characterized by limb darkening—a gradual decrease in specific intensity from the center of the stellar disk toward the limb. 
This effect becomes observationally significant in high-magnification events or during caustic crossings, where it introduces detectable deviations in the photometric light curve. 
Limb-darkening coefficients have been measured in a number of binary-lens microlensing events \citep{1999LD_Macho, 2001Albrow, 2003Fields}.
For events where the limb-darkening coefficient cannot be well-constrained directly from the data, tabulated values based on stellar atmospheric models are often adopted to compute magnifications \citep{2004Yoo, 2009Dong, 2010_binary_planet,fixedLD_Yee}.

For most cases, the single-parameter linear limb-darkening law is sufficient to accurately describe the features in the observed light curve. Its simplicity offers significant advantages: it is computationally efficient and introduces minimal degeneracy with other model parameters, which is crucial for robust and stable fitting procedures. More sophisticated parameterized models such as the square-root, quadratic, or logarithmic are sometimes employed to better capture the intensity drop-off near the limb \citep{2000Claret, An2002, 2007Heyrovsky}. For individual events with exceptional data quality, a non-parametric limb-darkening profile can be directly constrained \citep{2020FEM}. 

Several methods have been developed for calculating limb-darkened magnification.
For single-lens systems, general approaches were formulated by \citet{1997Heyrovsky, 2003Heyrovsky} and later by \citet{2019Witt}, though these are not directly applicable to more complex lens configurations.
\response{For binary lens, \cite{1997ApJ...477..580G} proposed using rings of constant surface brightness to incorporate the limb-darkened effects into contour integration. \cite{1998A&A...333L..79D} formulated a general boundary-integral method for limb-darkened sources. Building on \cite{eesunhong}, \cite{2010Bennett} developed an alternative method based on ray shooting.
\cite{2006ApJ...642..842D} developed a hybrid algorithm of contour integration and ray shooting.
\cite{VBBL2010} developed a widely-used contour-integration code based on concentric rings, whose algorithm includes an analytical high-order error estimate, allowing arbitrary-precision computations for limb-darkened sources.
}
The concentric-ring method samples the source with a series of concentric annuli and computes the limb-darkened magnification as the weighted sum of the magnifications of these annuli.
However, this approach has a fundamental limitation: its integration variable is essentially the image area of the concentric annuli, 
\revisionii{which is obtained only implicitly through the uniform-source magnification calculation.}
\revisionii{This prevents the algorithm from explicitly placing the quadrature nodes for integrand evaluation.
Consequently, it is difficult to employ high-order quadrature schemes—which require integrand evaluations at precisely controlled node locations—leading to a relatively slow convergence rate.
}

In this work, \refinement{we develop a new concentric-disk method for limb-darkening integration. Different from the concentric-ring method, our approach represents the limb-darkened source as a superposition of uniform disks with different radii and brightnesses. This construction corresponds to a reformulation of the integral that generates a new monotonic integrand and also permits pre-selected quadrature nodes and high-order numerical integration.} The integral expression becomes independent of the limb-darkening coefficient, enabling the reuse of the same integral term across multiple photometric bands and resulting in a significant gain in efficiency for multi-band analyses.

This paper is structured as follows. \S~\ref{sec:formula} derives the reformulated limb-darkening integral and establishes its mathematical properties. \S~\ref{sec:error} presents our numerical method, including error estimation schemes for both smooth regions and caustic crossings. \S~\ref{sec:result} demonstrates the algorithm's performance through systematic tests on a binary-lens source plane, comparing computational costs and convergence rates with conventional approaches.
\refinement{\S~\ref{sec:discussion} examines the validity of the error settings, extends the framework to astrometric calculations, and generalizes the method to more general limb-darkening laws.}

\section{Magnification Integration}\label{sec:formula}

For a limb-darkened source, the surface brightness $I$ at the center of the source star is higher than at the edge. The linear limb darkening effect, \revisionii{normalized to conserve the total source flux}, can be expressed as \citep{An2002}:
\begin{equation}
\label{eq:linear-limb-darkening}
\begin{aligned}
    I(r) &= \bar{I}f(r), \\ 
    f(r) &=  1-\Gamma + \frac{3\Gamma}{2} \sqrt{1-\frac{r^2}{R^2}},
\end{aligned}
\end{equation}
where $\bar{I}$ is the average surface brightness, and $R$ is the angular radius of the source. $r$ is the radial distance to the center, ranging from $0$ to $R$. Here $f$ is the profile function with the argument $r$. The linear limb-darkening coefficient $\Gamma$ ranges from $0$ to $1$, representing uniform brightness and zero brightness at the edge, respectively. 
The profile function $f(r)$ satisfies the normalization condition
\begin{equation}\label{eq:normalize}
    \int^R_0 2\pi r f(r) \mathrm{d}r = \int^R_0 2\pi r \mathrm{d}r = \pi R^2.
\end{equation}
There is another commonly used parameterization form as $f(r) \propto 1-a \left(1-\sqrt{1-r^2/R^2} \right)$, with $a$ ranging from $0$ to $1$. When the normalization condition in Equation~\eqref{eq:normalize} is satisfied, the two parameterizations are equivalent, related by the transformation $\Gamma = 2a/(3-a)$.

In microlensing events, the surface brightness is conserved during the lensing process. For a uniformly bright source, the total angular area of the images equals the magnification multiplied by the area of the source.
Defining the point-source magnification at different locations as $A_{\mathrm{point}}(r,\theta)$, where $r$ is the radial distance to the source center and $\theta$ is the azimuthal angle. The total angular area of the images $\Omega$ is given by:
\begin{equation}\label{eq:Area}
    \Omega(R) = \int^R_0 \left(\int^{2\pi}_0 A_{\mathrm{point}}(r,\theta) \mathrm{d}\theta \right)  r \mathrm{d}r = {A}_0(R)\pi R^2,
\end{equation}
where the ${A}_0(R)$ is the magnification of an uniformly bright extended source with a radius $R$. The subscript $0$ denotes the linear limb-darkening coefficient $\Gamma=0$, corresponding to uniform brightness.

For a limb-darkened source, the magnification ${A}_\Gamma$ is weighted by the surface brightness profile $f(r)$. Since $f(r)$ is independent of the azimuthal angle $\theta$, the expression can be simplified as
\begin{equation}\label{eq:M_Gamma}
\begin{aligned}
    {A}_\Gamma(R) &= \frac{1}{\pi R^2} \int^R_0 \left(\int^{2\pi}_0 A_{\mathrm{point}}(r,\theta) \mathrm{d}\theta \right) r f(r)\mathrm{d}r \\
    &= \frac{1}{\pi R^2}\int^R_0 \Omega'(r)f(r) \mathrm{d}r,
\end{aligned}
\end{equation}
where $\Omega'(r)$ denotes the derivative of angular area $\Omega(r)$ with respect to angular radius $r$. This integration form is equivalent to the concentric-ring method adopted in \citet{VBBL2010}.
\revisionii{We note that $\Omega(r)$ could be non-differentiable when the limb of a uniformly bright source touches a caustic, causing a jump discontinuity in the integrand of Equation~\eqref{eq:M_Gamma}. While this preserves mathematical convergence, it complicates numerical quadrature. This effect is further discussed and addressed in \S~\ref{sec:caustic_crossing_error}.}

The direct evaluation of the integrand in Equation~\eqref{eq:M_Gamma} is problematic for two reasons. First, the term $\Omega'(r)$ is not a native output of the contour integration method; since the algorithm is designed for computing the area $\Omega(r)$ directly, the derivative should be approximated afterward via numerical differentiation. This approximation substantially increases computational cost and introduces additional numerical error. Second, 
\revisionii{the derivative of the profile function $f(r)$ diverges as $r \to R$. 
Since numerical quadrature error estimates depend on derivatives of the integrand, this singular behavior results in larger numerical errors.}
Therefore, we \revisionii{apply integration by parts to Equation~\eqref{eq:M_Gamma}, yielding the reformulated expression:}
\begin{equation}\label{eq:area-integration}
\begin{aligned}
    {A}_\Gamma(R) &= \frac{1}{\pi R^2} \left[ \Omega(R)f(R) - \int^R_0\Omega(r)f'(r)dr \right]   \\
    &= {A}_0(R)f(R)+ \frac{1}{\pi R^2}\int^{f(0)}_{f(R)} \Omega\left[r(f)\right] \mathrm{d}f      \\
    &= {A}_0(R)f(R)+ \int^{f(0)}_{f(R)} {A}_0\left[r(f)\right]\frac{r^2(f)}{R^2} \mathrm{d}f.
\end{aligned}
\end{equation}

With mathematical transformations, the limb-darkened magnification ${A}_\Gamma(R)$ can be expressed as the sum of two components. The first component is the magnification for a uniform source. The second component, an integral term, can be reformulated with respect to the profile $f$ as the variable of integration. Since the profile function $f(r)$ is monotonic in $r$, its inverse function $r(f)$ can always be defined. 

Since the magnification at every point on the source is positive, the total imaging area, $\Omega(r)$, is a monotonically increasing function. Therefore, the total imaging area $\Omega[r(f)]$, corresponding to the integrand in Equation~\eqref{eq:area-integration}, decreases monotonically with respect to the integration argument $f$. This monotonic relationship enables a rigorous estimation of the integration error, which will be utilized in the subsequent analysis.

A useful change of integration variables is to treat the linear limb-darkening profile $f$ as a function of the direction cosine relative to the direction normal $\mu$:
\begin{equation}
\begin{aligned}
    &\mu = \sqrt{1-\frac{r^2}{R^2}},   \
    \mu_{r=0}=1, \mu_{r=R}=0,  \\
    &\mathrm{d}f = \frac{\mathrm{d}f}{\mathrm{d}\mu} \mathrm{d}\mu = 
    \frac{3\Gamma}{2} \mathrm{d}\mu. 
\end{aligned}
\end{equation}
The formulation of the linear limb-darkened magnification is simplified by adopting $\mu$ as the integration variable:
\begin{equation}\label{eq:LD_Inte}
\begin{aligned}
    {A}_1(R) &= \frac{3}{2}\int^1_0 \tilde{{A}}_0(\mu)(1-\mu^2)\mathrm{d}\mu,       \\
    {A}_\Gamma(R) &= (1-\Gamma){A}_0(R) + \Gamma {A}_1(R) ,
\end{aligned}
\end{equation}
where ${A}_1(R)$ denotes the magnification for the linear limb-darkening coefficient $\Gamma=1$. 
The tilde notation on $\tilde{A}_0$ indicates that it is treated as a function of the variable $\mu$. It is defined by the change of variable $\tilde{{A}}_0(\mu) = {A}_0\left[r(\mu)\right]$. In what follows, we refer to this integral term in ${A}_1$ as the linear limb-darkening integral.  This form reveals that the integral term is independent of the linear limb-darkening coefficient $\Gamma$. The computational benefit of this property is significant: for analyses involving multiple photometric bands with different limb-darkening coefficients, the same precomputed value of the integral can be reused, yielding a considerable efficiency gain.

The integrand of the linear limb-darkening integral $\tilde{{A}}_0(\mu)(1-\mu^2)$ is proportional to the image area $\Omega[r(\mu)]$ of the source with a radius $r(\mu)$. Due to the monotonicity of the integrand, along with the fact that the image area is always greater than the source area, the upper and lower bounds of the integral result can be estimated as follows:
\begin{equation}\label{eq:bound1}
\begin{aligned}
    (1-\mu^2) &\leq \tilde{{A}}_0(\mu)(1-\mu^2) &&\leq \tilde{{A}}_0(\mu=0) - \mu^2,  \\
    1 &\leq \quad \quad   {A}_\Gamma(R) &&\leq (1+\frac{\Gamma}{2}){A}_0 - \frac{\Gamma}{2}.
\end{aligned}
\end{equation}
Consequently, the magnification ${A}_\Gamma$ of a linearly limb-darkened source is bounded by ${A}_\Gamma < \frac{3}{2} {A}_0$, with ${A}_0$ denoting the uniform-source magnification.

\response{
We note that \cite{1998A&A...333L..79D} provides an alternative framework for computing limb-darkening magnifications via Green's theorem. 
\revisionii{For a given image region $\omega$ in the image plane with coordinates $(z_x, z_y)$, the magnification $A_\Gamma$ can be expressed as:
\begin{equation}\label{eq:uniform-image}
A_\Gamma=\frac{1}{\pi R^2}\int_\omega f(z_x,z_y){\rm d}z_x{\rm d}z_y,
\end{equation}
where $f(z_x,z_y)$ denotes the normalized source brightness profile mapped onto the image plane.
Green's theorem allows the integral to be rewritten as
\begin{equation}
\int_\omega
\left(
\frac{\partial Q}{\partial z_x}
-
\frac{\partial P}{\partial z_y}
\right)
{\rm d}z_x{\rm d}z_y
=
\oint_{\partial\omega}
P{\rm d}z_x
+
Q{\rm d}z_y .
\end{equation}
For the uniform source, $f=1$ in the image region $\omega$, and one possible choice is
$P=-z_y/2$ and $Q=z_x/2$.
For limb-darkened sources, \cite{1998A&A...333L..79D} constructed a different pair of functions, denoted here by $P_{\rm LD}$ and $Q_{\rm LD}$, such that the magnification can be obtained through a single contour integral.}
Instead of evaluating a sequence of uniform-source magnifications, that approach shifts the complexity to the contour integrands, namely the $P_{\rm LD}$ and $Q_{\rm LD}$ functions appearing in equations (19) and (20) of \cite{1998A&A...333L..79D}. \revisionii{Consequently, the numerical integration problem differs substantially from that of uniform-source contour integration.} 
In contrast, the concentric-disk method inherits the same contour-integration structure as the uniform-source problem, allowing the direct application of the high-order integration framework developed by \cite{VBBL2010}.
}

\section{Numerical Method and Error Estimation}\label{sec:error}

A fundamental limitation of the conventional integration formalism for the limb-darkening problem, which involves integrating over the image area of concentric annular rings, is its inability to 
\revisionii{evaluate the integrand at specific quadrature nodes.}
\revisionii{The reason is the integration variable is essentially the image area (Equation~\eqref{eq:M_Gamma}), and computing the integrand at a target node requires knowing the corresponding source radius. However, while the area is calculated from the radius, this mapping cannot be inverted to explicitly derive the radius from a given area. Consequently, the integrand values at these specific nodes remain inaccessible.}
\revisionii{This restriction confines traditional approaches to low-order quadrature schemes analogous to the trapezoidal rule, limiting their convergence rate.}
High-order numerical methods, such as Simpson's rule or Gaussian quadrature, explicitly require the deliberate placement of such nodes. In contrast, the integration form Equation~\eqref{eq:LD_Inte} adopts the profile function $\mu$ as the integration variable. Since the corresponding source radius $r$ for any given profile value $\mu$ can be computed explicitly, this reformulation enables the manual specification of optimal quadrature nodes, thereby facilitating the use of high-precision numerical integration schemes.

The error of the integral term in Equation~\eqref{eq:LD_Inte} could be split into two components: one generated by the difference between the computed magnification $\hat{{A}}_0(\mu)$ and the true magnification $\tilde{{A}}_0(\mu)$, and the other that arises from the error of the numerical integration itself. 
We denote the numerical integration result by $S$ and the associated error by $E$.
Applying the triangle inequality, we have
\begin{equation}
\begin{aligned}
    E =& \left| \int^1_0 \tilde{{A}}_0(\mu)(1-\mu^2) \, \mathrm{d}\mu - S \right| \\
    \leq& \left| \int^1_0 \left[\tilde{{A}}_0(\mu) - \hat{{A}}_0(\mu)\right](1-\mu^2) \, \mathrm{d}\mu \right|
        + \left| \int^1_0 \hat{{A}}_0(\mu)(1-\mu^2) \, \mathrm{d}\mu - S \right|. \\
\end{aligned}
\end{equation}
The two absolute-value terms on the right-hand side of the inequality correspond to the error from the uniform magnification computation (denoted $E_{\mathrm{{A}_0}}$) and the error from the numerical quadrature (denoted $E_{\mathrm{S}}$), respectively.

The error estimation for the uniform-source magnification is straightforward. Through simple mathematical transformations, we can derive the relationship between $E_{\mathrm{{A}_0}}$ and the tolerance $\mathrm{Tol}_{\mathrm{{A}_0}}$ used in the uniform-source computation:
\begin{equation}
    \begin{aligned}
        E_{\mathrm{{A}_0}} &= \left| \int^1_0 \left[\tilde{{A}}_0(\mu) - \hat{{A}}_0(\mu)\right](1-\mu^2) \, \mathrm{d}\mu \right| \\
        &<\max\left[\tilde{{A}}_0(\mu) - \hat{{A}}_0(\mu)\right] \int^1_0 (1-\mu^2) \, \mathrm{d}\mu     \\
        &< \frac{2}{3} \mathrm{Tol}_{\mathrm{{A}_0}}
    \end{aligned}
\end{equation}

Combining this result with Equation~\eqref{eq:LD_Inte}, we obtain the final upper bound for the error of the linearly limb-darkened source magnification as:
\begin{equation}
\begin{aligned}
    E&<(1-\frac{\Gamma}{3}) \mathrm{Tol}_{\mathrm{{A}_0}} + \Gamma \mathrm{Tol}_{\mathrm{S}}   \\
    &< \mathrm{Tol}_{\mathrm{{A}_0}} + \Gamma \mathrm{Tol}_{\mathrm{S}} .
\end{aligned}
\end{equation}
In practice, the tolerance for the uniform-source magnification, ${\rm Tol}_{{A}_0}$, is set smaller than the final limb-darkened magnification tolerance, ${\rm Tol}_{{A}_0} = {\rm Tol}_{\rm {A}_\Gamma}/3 $. The tolerance for the numerical integration, ${\rm Tol}_{\rm S}$, is taken as ${\rm min}\left(1, {\rm Tol}_{{A}_\Gamma} / \Gamma\right)$.

\subsection{Simpson Integration and Error}

The numerical integration scheme employed in this study is the Adaptive Simpson's Method. The procedure begins by computing a Simpson integral over the entire integration interval. This parent interval is then bisected, and the Simpson integration is performed separately on the left subinterval and the right subinterval. This subdivision process is applied recursively until the estimated error falls below a preset tolerance.

To apply Simpson's rule over an integration interval $[\mu_{\mathrm{L}},\mu_{\mathrm{R}}]$, the integrand must be evaluated at the endpoints and the midpoint. Given the integrand $\alpha(\mu):=\tilde{{A}}_0(\mu)(1-\mu^2)$, the integral estimate and its associated error are given by:
\begin{equation}\label{eq:Simpson}
    \begin{aligned}
         \left[ \int^{\mu_{\mathrm{R}}}_{\mu_{\mathrm{L}}}\alpha(\mu)\mathrm{d}\mu \right]_{\mathrm{Sp}} 
        &= \frac{\mu_{\mathrm{R}}-\mu_{\mathrm{L}}}{6}\left[ \alpha\left(\mu_{\mathrm{L}}\right) + 4\alpha\left(\frac{\mu_{\mathrm{L}}+\mu_{\mathrm{R}}}{2}\right) + \alpha\left(\mu_{\mathrm{R}}\right)\right],   \\
        E_{\mathrm{Sp}} &= -\frac{(\mu_{\mathrm{R}}-\mu_{\mathrm{L}})^5}{2880} \alpha^{(4)}(\xi),
    \end{aligned}
\end{equation}
where $\xi \in [\mu_{\mathrm{L}},\mu_{\mathrm{R}}]$ is an intermediate point guaranteed by the mean value theorem, and $\alpha^{(4)}$ denotes the fourth-order derivative.

Simpson's rule possesses third-order algebraic precision, meaning it yields exact results for polynomials of degree up to three. 
In the context of the linear limb-darkening integral, the constant and linear components of the magnification $\tilde{{A}}_0$ are integrated exactly; the leading-order integration error arises from the second and higher derivatives of $\tilde{{A}}_0(\mu)$. This property significantly accelerates the convergence of the numerical integration.

Since the fourth derivative of the linear limb-darkening integral cannot be directly computed, we estimate the integration error by comparing the integral over a parent interval with the sum of the integrals over its two subintervals. Specifically, for a parent interval $(\mu_{\mathrm{L}}, \mu_{\mathrm{R}})$ with midpoint $\mu_{\mathrm{C}}$, we define the two subintervals $(\mu_{\mathrm{L}}, \mu_{\mathrm{C}})$ and $(\mu_{\mathrm{C}}, \mu_{\mathrm{R}})$, whose midpoints are denoted as $\mu_{\mathrm{LC}}$ and $\mu_{\mathrm{RC}}$, respectively. The  normal error item $E_{\mathrm{n}}$ is estimated as:
\begin{equation}
\begin{aligned}
E_{\mathrm{n}} &= \biggl| \biggl[ \int^{\mu_{\mathrm{R}}}_{\mu_{\mathrm{L}}}\alpha(\mu)\mathrm{d}\mu \biggr]_{\mathrm{Sp}} 
      - \biggl[ \int^{\mu_{\mathrm{C}}}_{\mu_{\mathrm{L}}}\alpha(\mu)\mathrm{d}\mu \biggr]_{\mathrm{Sp}}   - \biggl[ \int^{\mu_{\mathrm{R}}}_{\mu_{\mathrm{C}}}\alpha(\mu)\mathrm{d}\mu \biggr]_{\mathrm{Sp}}  \biggr| \\
    &= \frac{\mu_{\mathrm{R}}-\mu_{\mathrm{L}}}{12} \bigl| \alpha(\mu_{\mathrm{L}}) - 4\alpha(\mu_{\mathrm{LC}}) 
       + 6\alpha(\mu_{\mathrm{C}}) 
     - 4\alpha(\mu_{\mathrm{RC}}) + \alpha(\mu_{\mathrm{R}}) \bigr|.
\end{aligned}
\end{equation}
This error expression is related to a fourth-order finite difference approximation of the derivative. The standard fourth-order central difference approximation at the midpoint $\mu_{\mathrm{C}}$ takes the form:
\begin{equation}\label{eq:4th-derivative}
\begin{aligned}
    &\alpha^{[4]}(\mu_{\mathrm{C}})  \approx\frac{\alpha(\mu_{\mathrm{L}}) - 4\alpha(\mu_{\mathrm{LC}}) + 6\alpha(\mu_{\mathrm{C}}) - 4\alpha(\mu_{\mathrm{RC}}) + \alpha(\mu_{\mathrm{R}})}{(\mu_{\mathrm{R}}-\mu_{\mathrm{L}})^4 \ / \ 256}.
\end{aligned}
\end{equation}
Here, the upper square brackets indicate that $\alpha^{[4]}$ is obtained through numerical differentiation rather than analytical computation. Accordingly, for normal cases, the error estimate $E_{\mathrm{n}}$ effectively provides a numerical approximation of the fourth derivative at the midpoint $\mu_{\mathrm{C}}$:
\begin{equation}
\begin{aligned}
    E_{\mathrm{n}} = \frac{(\mu_{\mathrm{R}}-\mu_{\mathrm{L}})^5}{3072}\left| \alpha^{[4]}(\mu_{\mathrm{C}}) \right|,
\end{aligned}
\end{equation}
which aligns with the theoretical error for Simpson's rule in Equation~\eqref{eq:Simpson}.

\subsection{Caustic-crossing Error}\label{sec:caustic_crossing_error}
In microlensing events, 
\revisionii{caustics refer to the source-plane locations where the point-source magnification becomes infinite. Their counterparts in the image plane are the critical curves, which map onto the caustics through the lens equation. Therefore, finite-source effects are most pronounced in the vicinity of caustics.}
When a source traverses a caustic, the first-order derivative of the magnification could change abruptly. This leads to a discontinuity in the fourth-order derivative of the linear limb-darkening integrand, which undermines the reliability of the standard error estimate.
\revisionii{A potential solution is to break the integration interval at these non-differentiable points. However, such a piecewise strategy requires determining the exact $\mu$ values where the source limb touches the caustics. This task is equivalent to finding the minimum distance from the caustic to the source center, which is generally computationally expensive due to the complexity of the caustic.}
\revisionii{Instead of locating these discontinuities explicitly, we adopt a more practical error-control strategy.}
The \texttt{Twinkle} algorithm provides the number of caustic crossings for the source by counting the number of images crossing the critical curve. When adjacent $\mu$ values correspond to different numbers of caustic crossings, we introduce an additional error term $E_c$ to account for the resulting discontinuity.

The form of $E_c$ is derived from an estimation of the integral's upper and lower bounds. Owing to the monotonically decreasing nature of the linear limb-darkening integrand, the integral over the interval $(L,R)$ can be bounded from above and below using the left and right rectangle rules, respectively, evaluated at the quadrature nodes. For the set of equidistant nodes $(\mu_{\mathrm{L}}, \mu_{\mathrm{LC}}, \mu_{\mathrm{C}}, \mu_{\mathrm{RC}}, \mu_{\mathrm{R}})$, the upper bound $\overline{S}_{\mathrm{LR}}$ and the lower bound $\underline{S}_{\mathrm{LR}}$ of the integral are given as:
\begin{equation}\label{eq:ul_bounds}
    \begin{aligned}
        \overline{S}_{\mathrm{LR}} &= \frac{\mu_{\mathrm{R}}-\mu_{\mathrm{L}}}{4} \left[ \alpha(\mu_{\mathrm{L}}) +\alpha(\mu_{\mathrm{LC}}) + \alpha(\mu_{\mathrm{C}})+\alpha(\mu_{\mathrm{RC}}) \right],    \\
        \underline{S}_{\mathrm{LR}} &= \frac{\mu_{\mathrm{R}}-\mu_{\mathrm{L}}}{4} \left[ \alpha(\mu_{\mathrm{LC}}) +\alpha(\mu_{C}) + \alpha(\mu_{\mathrm{RC}})+\alpha(\mu_{\mathrm{R}}) \right].
    \end{aligned}
\end{equation}
Thus, analogous to Equation~\eqref{eq:bound1}, we derive a relatively loose lower bound for the linear limb-darkening integral. This is expressed as:
\begin{equation}
    \begin{aligned}
        \int^1_0 \tilde{{A}}_0(\mu)(1-\mu^2) \mathrm{d}\mu >  \tilde{{A}}_0(\mu_0) \left( \mu_0 - \mu_0^3\right),
        \quad
        \text{for } \mu_0 \in [0, 1].&
    \end{aligned}
\end{equation}

The bounds provided by Equation~\eqref{eq:ul_bounds} depend only on monotonicity and are independent of any other properties, such as continuity. Consequently, while these estimates are highly robust, they are also relatively conservative. In practice, a coefficient $C_{\mathrm{cross}}$ is adopted for error estimation, the default value of which is $ C_{\mathrm{cross}} = 0.1$:
\begin{equation}\label{eq:error_cross}
    E_{\mathrm{c}} = C_{\mathrm{cross}}(\overline{S}_{\mathrm{LR}} - \underline{S}_{\mathrm{LR}}).
\end{equation}

\subsection{Hidden Caustic Error}

Using changes in the number of caustic crossings as the sole criterion for adding an extra error term can be dangerous. For instance, when the source is large relative to a small caustic, the caustic may be entirely hidden between two adjacent sampled $\mu_{\mathrm{L}}$ and $\mu_{\mathrm{R}}$, without being detected by a crossing count. Therefore, we implement a geometric criterion that checks the positions of reference caustic points.
 
\revisionii{In gravitational microlensing, the lens equation describes the mapping from the image position $z$ to the source position $\zeta$. Accordingly, the caustics $\zeta_c$ in the source plane are mapped from the corresponding critical curves $z_c$ in the image plane through this equation. For a binary lens, the critical curves are mathematically determined by solving a set of quartic equations with parameter $\psi$ \citep{Witt1990}:}
\begin{equation}\label{eq:critical_curve_sampling}
    \frac{m_1}{(z_c-z_1)^2} + \frac{m_2}{(z_c-z_2)^2} = \mathrm{e}^{\mathrm{i}\psi},\ \ \psi \in [0,2\pi)
\end{equation}
\revisionii{The caustics are then derived by substituting these critical curves into the binary lens equation:}
\begin{equation}\label{eq:caustic_sampling}
    \zeta_c = z_c + \frac{m_1}{\bar{z}_1-\bar{z}_c} + \frac{m_2}{\bar{z}_2-\bar{z}_c},
\end{equation}
\revisionii{where $z_1$ and $z_2$ denote the lens positions in units of the Einstein radius. All lengths appearing in this work are expressed in units of the Einstein radius, the natural length scale in microlensing. The quantities $m_1$ and $m_2$ are their normalized masses $(m_1 + m_2 =1)$. The overbar (e.g., $\bar{z}$) denotes a complex conjugate.}
By selecting a specific parameter $\mathrm{e}^{\mathrm{i}\psi}=-1$, we can solve for four reference points that lie on the caustic structure. Their configuration depends on the lens geometry, \revisionii{specifically the binary separation $s$, defined by the separation between the two lens components, $|z_1-z_2|$}. 
\revisionii{For wide ($s\gg1$) and close ($s\ll1$) binaries, the caustics split into a central (stellar) caustic located near the host star and one (wide) or two (close) planetary caustics located away from the host star. Near $s\approx1$, these structures merge into a single resonant caustic.}
For $s \ll 1$, two points lie on the left and right cusps of the central \revisionii{(stellar)} caustic,  and the other two lie on the two detached planetary caustics, respectively; for $s\approx1$, all four lie on the resonant caustic; for $s\gg1$, they correspond to the left and right cusps of the central and planetary caustics.

For adjacent $\mu_{\mathrm{L}}$ and $\mu_{\mathrm{R}}$, we check if any of the four caustic reference points are located inside the source disk at the larger radius $\mu_{\mathrm{L}}$ but outside at the smaller radius $\mu_{\mathrm{R}}$. If such a point exists and both $\mu_{\mathrm{L}}$ and $\mu_{\mathrm{R}}$ report zero caustic crossings, we conclude that a caustic is hidden between them and add an error term $E_{\mathrm{h}}$ similar to $E_{\mathrm{c}}$ to this interval.

\begin{equation}\label{eq:error_hidden}
    E_{\mathrm{h}} = \overline{S}_{\mathrm{LR}} - \underline{S}_{\mathrm{LR}}.
\end{equation}

The hidden caustic error approach is equally applicable to more complex lens systems, requiring only modification of the related equations ~\eqref{eq:critical_curve_sampling} and ~\eqref{eq:caustic_sampling}. In the case of a single lens system, the caustic degenerates to the coordinate origin.

\subsection{Initial Step Error}
The error estimation described above requires at least five quadrature nodes to compute the fourth-order derivative ~\ref{eq:4th-derivative}. However, in regions where the magnification varies only weakly with radius, the integration error may already fall below the tolerance with only three nodes. In such cases, the computation of the two additional nodes is redundant. This situation frequently occurs when the tolerance is relatively loose or the source is far from caustics. To improve efficiency, we construct a special error estimator for the initial step that uses only three nodes $\mu_0=0,\mu_{\frac{1}{2}}=\frac{1}{2},\mu_1=1$. 
This estimator mainly relies on the second-order derivative of the magnification and employs a small empirical coefficient $C_{\mathrm{ini}} = 0.1$ to accelerate convergence while maintaining a reliable error bound:
\begin{equation}\label{eq:error_initial}
    \begin{aligned}
        E_2 &=  \left| \tilde{{A}}_0(\mu_0) -2\tilde{{A}}_0(\mu_{\frac{1}{2}}) +\tilde{{A}}_0(\mu_1) \right|\\
        E_1 &= \left| \tilde{{A}}_0(\mu_0) -\tilde{{A}}_0(\mu_1) \right|\\
        E_{\mathrm{ini}} &= C_{\mathrm{ini}} \ \mathrm{max} \left\{ E_2, \frac{1}{3}E_1\right\}.
    \end{aligned}
\end{equation}
The $\tilde{{A}}_0(\mu_1)$ corresponds to the point-source magnification, which is computationally much cheaper than the extended source magnification. Consequently, this refinement allows the limb-darkening integration to converge with as few as two extended source magnification evaluations, significantly reducing the computational cost in smooth regions. The initial step error is used only at the beginning, acting as a temporary substitute for the standard Simpson error and has no impact on the error assessment in subsequent recursive steps.

\subsection{Relative Error}
The error mentioned above refers to the absolute error. For cases where a relative error criterion is required, the lower bound established by monotonicity in Equation~\eqref{eq:ul_bounds} is used as the denominator to compute an upper bound for the relative error, which subsequently serves as the criterion for integration convergence. This approach avoids underestimating the relative error due to an excessively large numerical integral value. 
It is particularly effective in the early iterations at reducing the risk of premature termination, thereby ensuring more robust convergence behavior.

\section{Result}\label{sec:result}

\subsection{ Source Plane}
Utilizing the integration method and error estimation framework described above, we have constructed a new limb-darkening integration algorithm for \texttt{Twinkle} \response{\citep{Twinkle_Zenodo}}. To validate the performance of this implementation, we evaluate the computational cost of calculating limb-darkening effects for different microlensing systems.
As shown in Equation~\eqref{eq:LD_Inte}, the magnification for a linearly limb-darkened source is a linear combination of the uniform-source magnification and the magnification for a linear limb-darkening coefficient $\Gamma=1$. Therefore, in this test, we consider only the case $\Gamma=1$. The relative tolerance for the calculation is set to $10^{-6}$.

\begin{figure*}
\centering
\begin{subfigure}{0.94\textwidth}
    \centering
    \includegraphics[width=0.94\textwidth]{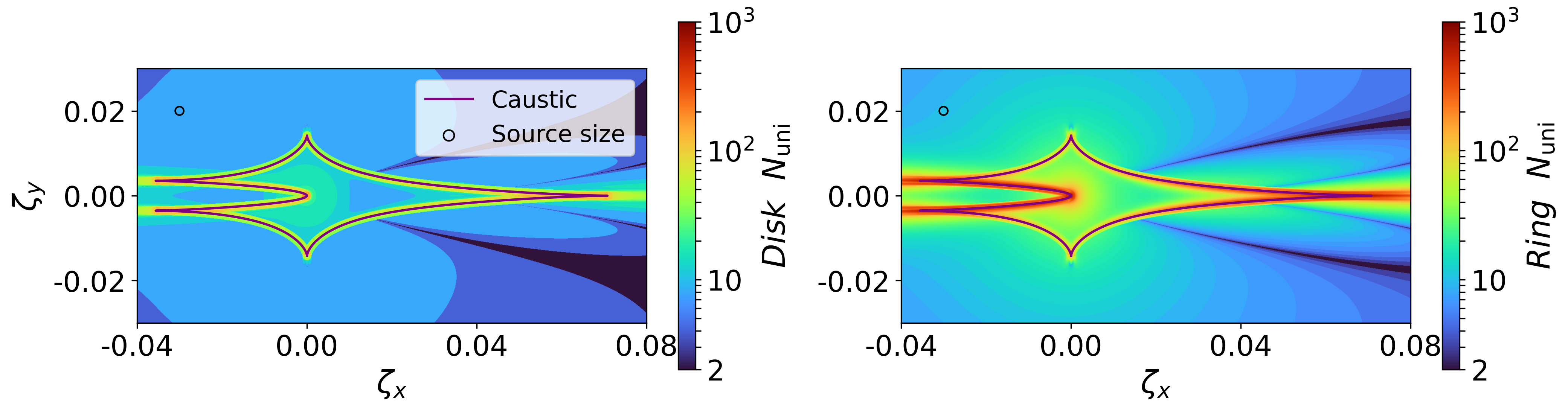}
    \caption{Binary separation $s=1$, mass ratio $q=0.0001$, source radius $\rho=0.001$.}
    \label{fig:2D_comparision_a}
\end{subfigure}

\begin{subfigure}{0.94\textwidth}
    \centering
    \includegraphics[width=0.94\textwidth]{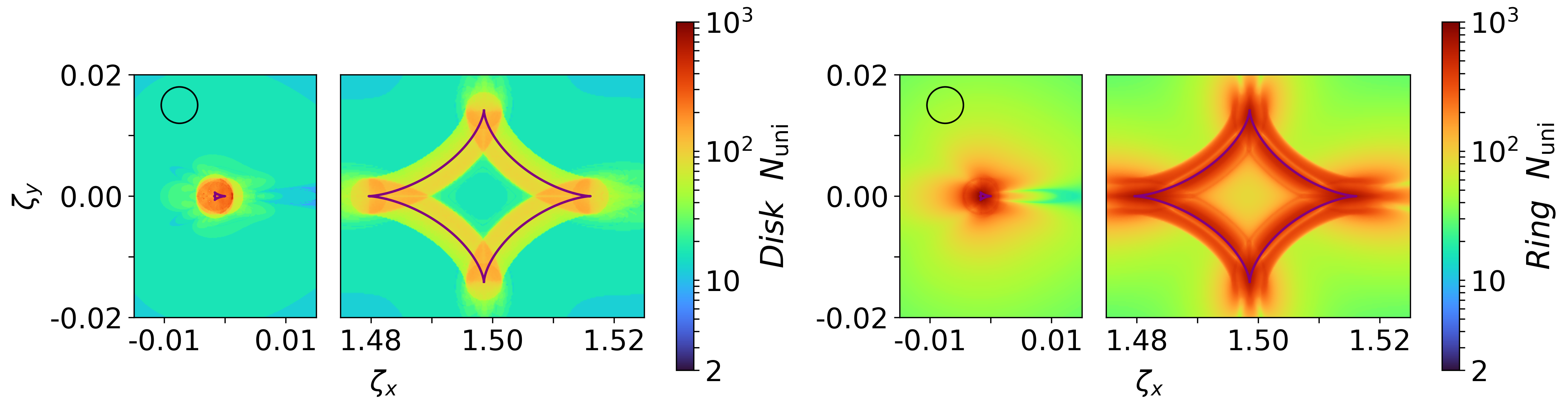}
    \caption{Binary separation $s=2$, mass ratio $q=0.001$, source radius $\rho=0.003$.}
    \label{fig:2D_comparision_b}
\end{subfigure}

\caption{Comparison of computational performance between \texttt{Twinkle}'s concentric-disk limb-darkening integration method (Equation~\eqref{eq:LD_Inte} with adaptive Simpson integration) and the concentric-ring method \citep{VBBL2010} \revisionii{in the source plane ($\zeta_x-\zeta_y$)}.
\revisionii{The labels "Disk" and "Ring" refer to the concentric-disk and concentric-ring methods, respectively.}
The relative tolerance is set to $10^{-6}$. Purple curves denote the caustics. 
Source sizes are indicated by the circle in the upper-left corner of each subfigure. In subfigure (b) the horizontal axis is split into two segments, $(-0.015,0.015)$ and $(1.475,1.525)$, in order to display both the stellar and planetary caustics clearly.
The colors in each panel represent the number of uniform-source magnification evaluations $N_{\rm{uni}}$ required for convergence, plotted on a logarithmic color scale (identical for both panels). 
}

\label{fig:2D_comparision}
\end{figure*}

Figure~\ref{fig:2D_comparision} demonstrates the computational performance of the concentric-disk method compared with the concentric-ring approach. 
\revisionii{To illustrate the performance of the algorithm across different caustic topologies and source radius, } subfigure~\ref{fig:2D_comparision_a} corresponds to a small source and a large resonant caustic 
($s=1,q=0.0001,\rho = 0.001$\revisionii{, where $s$ is the binary separation, $q=m_2/m_1$ is the mass ratio, and $\rho$ is the source radius in units of the Einstein radius.}), while subfigure~\ref{fig:2D_comparision_b} adopts a larger source together with a small central caustic and a relatively large planetary caustic ($s=2,q=0.001,\rho = 0.003$). 
In each subfigure, the horizontal \revisionii{$\zeta_x$} and vertical axes \revisionii{$\zeta_y$} represent positions on the source plane. The purple curves are the caustics, and the relative size of the source is indicated by the circle in the upper-left corner.

\revisionii{
The characteristic low-cost and high-cost regions in Figure~\ref{fig:2D_comparision} can be understood directly from Equation~\eqref{eq:LD_Inte}.
The limb-darkened magnification differs from the uniform-source magnification only because the uniform-source magnification $\tilde{A}_0(\mu)$ varies with source radius. 
When uniform-source magnification shows only weak dependence on direction cosine $\mu$, $\tilde{A}_0(\mu) \approx A_0$, the limb-darkened magnification naturally approaches the uniform-source limit $A_1 \to A_0$. 
Away from caustics, the point-source magnification field is smooth and admits the quadrupole approximation \citep{2018MNRAS.479.5157B, 2009ApJ...690.1772P, Gould2008},
\begin{equation}\label{eq:quadrupole}
    A_0(r) = A_{\rm{point}} + \frac{r^2}{8}\Delta A_{\rm{point}} + o\left(r^2\right),
\end{equation}
where $\Delta=\partial^2_x + \partial^2_y$ is the Laplacian operator.
In regions where the quadrupole term is very small, the uniform-source magnification is insensitive to the source radius, and the integrand in Equation~\eqref{eq:LD_Inte} approaches a quadratic function $A_\mathrm{point}(1-\mu^2)$. Since Simpson integration is exact for quadratic functions, only a very small number of quadrature nodes are required to reach the target accuracy. 
By contrast, near caustics the point-source magnification field becomes non-differentiable and the quadrupole approximation breaks down. The uniform-source magnification then becomes much more sensitive to source radius, 
requiring much more uniform sources to achieve convergence.
This explains both the dark low-cost regions and the bright high-cost structures that trace the caustics in Figure~\ref{fig:2D_comparision}. 
In addition, the concentric-disk method systematically requires fewer evaluations than the concentric-ring method throughout the source plane, reflecting the advantage of its higher-order numerical integration scheme.
}

In subfigure~\ref{fig:2D_comparision_a}, two dark ``wings'' appear to the right of the caustic.
\revisionii{
These wings correspond to regions where the quadrupole correction is particularly small, 
so that the uniform-source magnification remains nearly constant over the range of radii spanned by the source.
Consequently, the limb-darkening integrand is close to a quadratic function, and the integration converges with as few as two uniform-source magnifications.
}
This is precisely the regime for which the initial-step error estimator (Equation~\eqref{eq:error_initial}) was designed.
Away from caustic crossings, the concentric-disk method typically achieves a relative accuracy of $10^{-6}$ using no more than 8 uniform-source magnifications (often 2, 4, or 8), while the concentric-ring approach frequently requires several dozen magnifications to converge, corresponding to a computational cost of about $60\%$ that of the concentric-ring method.
For source positions that cross a caustic, the concentric-disk method needs roughly $10^1$ uniform magnifications, while the concentric-ring method often demands over $10^2$, yielding a reduction in cost to $\lesssim20\%$.
The largest values of $N_{\rm{uni}}$ for both methods occur near $(0,0)$, the region of highest magnification. There, the concentric-disk method converges with about $200$ uniform magnifications, compared to approximately $900$ required by the concentric-ring method.

In subfigure~\ref{fig:2D_comparision_b}, the source is larger, with a size comparable to that of the planetary caustic and exceeding the scale of the central (stellar) caustic.
This results in a stronger radial difference in magnification across the source disk.
Consequently, a greater number of uniform-source magnification evaluations (typically $\gtrsim30$) is required to achieve convergence.
Because the concentric-disk method increases the algebraic order of numerical integration, the acceleration becomes more significant here, requiring less than $30\%$ of the computational cost of the concentric-ring approach in most regions.

\subsection{ Convergence Behavior }\label{sec:converge_behavior}

\begin{figure*}
\centering
\includegraphics[width=\textwidth]{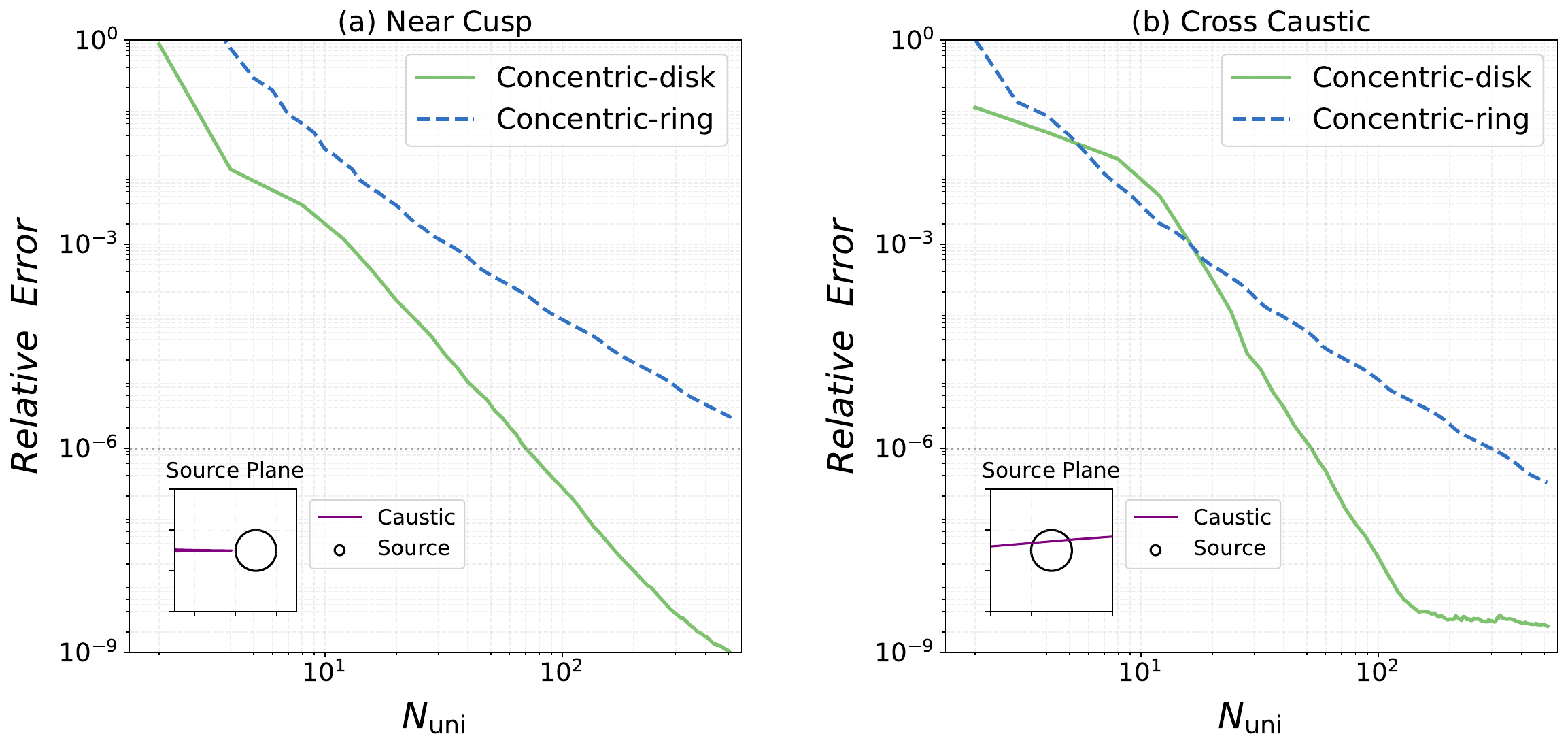}

\caption{
Convergence behavior of relative error as a function of uniform-source magnification evaluations $N_{\mathrm{uni}}$ for two numerically challenging source positions.
The lens parameters are $s=1$, $q=10^{-4}$ and the source radius is $\rho=0.001$.
The gray horizontal line indicates the tolerance ($10^{-6}$) used for both the uniform-source computation and the limb-darkening integration.
Panel (a) shows a source centered at $(0.072,0)$ just outside the right cusp of the resonant caustic; panel (b) shows a source at $(0.04,-0.002)$ crossing a smooth (non-cusp) caustic segment (insets show source positions relative to caustics).
The blue dashed lines are the relative error of the concentric-ring method, while the green solid lines represent the relative error of \texttt{Twinkle}'s concentric-disk method. 
Power-law fits in the range $20 < N_{\mathrm{uni}} < 100$ yield convergence exponents of $-4.80$ (\texttt{Twinkle}'s new method) and $-2.32$ (concentric-ring method) for the near-cusp case, and $-5.52$ and $-2.38$ respectively for the caustic-crossing case. These values confirm the concentric-disk method's faster-than-$N^{-4}$ scaling versus the concentric-ring $N^{-2}$ scaling, particularly effective when the source crosses the caustic.
For $N_{\mathrm{uni}} > 100$, errors plateau at $\gtrsim 10^{-9}$ due to finite precision in the uniform-source magnification computation.}

\label{fig:error_decrease}
\end{figure*}

Figure~\ref{fig:error_decrease} shows the convergence behavior of the relative error as the number of uniform-source magnification evaluations $N_{\rm{uni}}$ increases, focusing on two representative positions that have high computational cost in Figure~\ref{fig:2D_comparision_a}. The lens parameters are the same as $s=1$ and $q=10^{-4}$, and the source radius is $\rho=0.001$. Panel (a) illustrates the error evolution when the source is close to a cusp, with the source center at $(0.072,0)$. Panel (b) shows the case where the source crosses a regular (non-cusp) caustic, with the source center at $(0.04,-0.002)$. In both panels, the horizontal axis is $N_{\rm{uni}}$ and the vertical axis is the relative error estimated at each integration step; both axes use a logarithmic scale. The gray horizontal line at $10^{-6}$ indicates the tolerance set for the uniform-source magnification computation. The blue dashed line represents the relative error of the concentric-ring method, while the green solid line corresponds to the relative error of \texttt{Twinkle} concentric-disk method. 
An inset in the lower-left corner of each panel displays the position of the source relative to the caustic. In panel (a), the source lies just outside the right cusp of the resonant caustic. In panel (b), the source crosses a smooth part of a caustic where no cusp is present.

The errors in both panels decrease approximately in a straight line on the logarithmic axes, indicating power-law convergence with increasing $N_{\rm{uni}}$. We fit the power-law exponents over the interval $20 < N_{\rm{uni}} < 100$. For the near-cusp case (panel a), the \revisionii{fitted} exponents \revisionii{for this specific configuration} are $-2.32$ for the concentric-ring error and $-4.80$ for concentric-disk error. These values agree well with theoretical expectations: 
on uniform integration nodes, the concentric-ring and concentric-disk methods would scale as $N^{-2}_{\rm{uni}}$ and $N^{-4}_{\rm{uni}}$, respectively. Because both methods employ adaptive sampling that preferentially refines the highest-error intervals, their observed convergence is somewhat faster than the uniform-grid rates. 
In the caustic-crossing case (panel b), the fitted exponents are about 
$-2.38$ for the concentric-ring error and $-5.52$ for concentric-disk error. When the source crosses a caustic, the dominant error arises when the source contour is tangent to the caustic. The adaptive strategy is particularly effective in such regimes, allowing the concentric-disk method to achieve a convergence rate even steeper than $N^{-5}_{\rm{uni}}$.

The power-law fit is restricted to $N_{\rm{uni}} < 100$ because, beyond this value, the integration error falls far below the tolerance of the uniform-source magnification computation. At larger $N_{\rm{uni}}$, the error is consequently dominated by the noise in the uniform-source magnifications, and thus deviates from the ideal scaling of the integration scheme. This effect is clearly visible in panel b for $N_{\rm{uni}} > 100$, where the integration error reaches a floor greater than $10^{-9}$ and does not decrease further.
Besides, as shown in panel b, the error of the concentric-disk method can temporarily exceed that of the concentric-ring method during the first few iterations. Such occurrences are rare and confined to the early stage of the computation. Therefore, this phenomenon has little influence on the overall enhancement in computational efficiency.

\subsection{Computation Efficiency}

\begin{figure*}
\centering
\includegraphics[width=\textwidth]{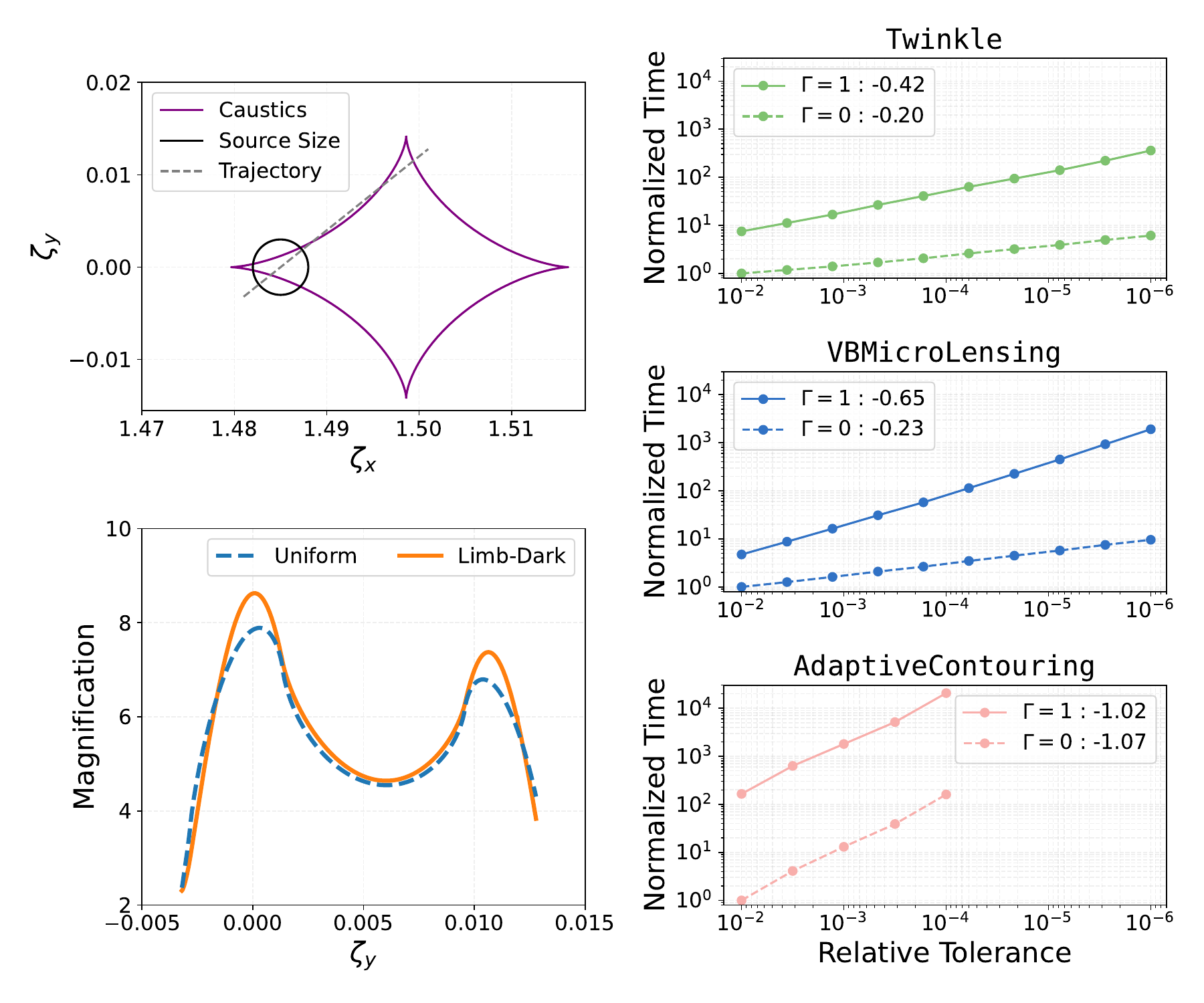}

\caption{
\response{
Computational time performance for a caustic-crossing light curve. Top left: Microlensing system configuration. The black circle represents the source crossing the purple caustic along the gray trajectory. The lens parameters are identical to those in Figure~\ref{fig:2D_comparision_b}, with $\rm{tan}(\alpha) =0.8$, where $\alpha$ is the trajectory position angle. Bottom left: Magnification curves. The blue dashed line corresponds to a uniform-brightness source, and the orange solid line represents a linear limb-darkened source with $\Gamma=1$. Right panels: Normalized computational time as a function of the \revisionii{relative tolerance} for different limb-darkening integration algorithms. Solid lines denote uniform-brightness sources, while dashed lines indicate limb-darkened sources. The scaling of computational time with accuracy follows an approximate power law for all methods. The green lines represent \texttt{Twinkle}, with slopes of $-0.20 (\Gamma=0)$ and $-0.42 (\Gamma=1)$. The blue lines correspond to \texttt{VBMicroLensing}, with slopes of $-0.23$ and $-0.65$. The pink lines show the \texttt{AdaptiveContouring} method, with slopes of $-1.07$ and $-1.02$.
}
}

\label{fig:time_compare}
\end{figure*}

\response{
We evaluate the computational efficiency of our limb-darkening method, implemented in \texttt{Twinkle}. We also compare it with the concentric-ring approach realized in \texttt{VBMicroLensing} \citep{VBML2024}, as well as the method of \citep{1998A&A...333L..79D} as incorporated into \texttt{AdaptiveContouring} \citep{AC_2007MNRAS.377.1679D}.
\texttt{Twinkle} is GPU-accelerated (run on an NVIDIA RTX 4090 at 2.2 GHz), whereas \texttt{VBMicroLensing} and \texttt{AdaptiveContouring} are CPU-based (run on an AMD EPYC 7H12 at 2.6 GHz). The versions of \texttt{VBMicroLensing} and \texttt{AdaptiveContouring} come from \texttt{MulensModel} \citep{MulensModel_2019A&C....26...35P, MulensModel_2018ascl.soft03006P}. 
\revisionii{It is important to note that the \texttt{AdaptiveContouring} version tested here is a specific implementation of the adaptive contouring method described by \cite{AC_2007MNRAS.377.1679D}, without subsequent optimizations on either the refinement of the image contour or the interpolation scheme. 
Furthermore, \cite{AC_2007MNRAS.377.1679D} is a special choice of incorporating the broader principles introduced by \cite{1998A&A...333L..79D}, rather than the only possible realization. Therefore, the results presented here should be interpreted as a comparison with this specific implementation rather than a general assessment of the Green's-theorem approach.}
In the following, we normalize computation times to the uniform-brightness case to assess their comparative performance for limb-darkening sources.
}

\response{
Figure~\ref{fig:time_compare} illustrates the computational efficiency of different limb-darkening algorithms. 
The lens system configuration is identical to that shown in Figure~\ref{fig:2D_comparision_b}, with a binary separation $s=2$, mass ratio $q=0.001$, and source radius $\rho=0.003$. The source trajectory crosses the planetary caustic, intersecting the x-axis at $(1.485,0)$ with the tangent of position angle $ \tan(\alpha)=0.8$. As detailed in the top-left panel of Figure~\ref{fig:time_compare}, the trajectory continuously intersects the caustic, representing a regime where limb-darkening effects are most significant. We compute magnifications for both a uniform-brightness source and a linearly limb-darkened source ($\Gamma=1$), as shown in the bottom-left panel. Due to the caustic crossing, the light curve exhibits characteristic U-shaped features, with the limb-darkened source (orange solid line) reaching a higher peak magnification during the crossing compared to the uniform source (blue dashed line).}

\response{The right panels of Figure~\ref{fig:time_compare} display the relationship between computational time and \revisionii{the input relative tolerance (the accuracy control parameter).} For all three methods, the computation times are normalized relative to the cost of calculating the uniform-brightness source at \revisionii{an input tolerance} of $10^{-2}$. Solid lines represent the uniform-source calculations, while dashed lines correspond to the limb-darkened calculations. The computational time for each algorithm scales with accuracy as an approximate power law, but the exponents vary significantly. For \texttt{Twinkle} (green lines), the power-law exponents are $-0.20$ for the uniform source and $-0.42$ for the limb-darkened source. \texttt{VBMicroLensing} (blue lines) yields exponents of $-0.23$ and $-0.65$, respectively. While \texttt{AdaptiveContouring} (pink lines) shows very similar exponents for both cases: $-1.07$ for the uniform source and $-1.02$ for the limb-darkened source.}
\revisionii{Regarding absolute computational performance, for the present caustic-crossing light curve, the uniform-source calculation of \texttt{Twinkle} on a single GPU computes roughly as fast as $\gtrsim50$ single CPU threads running \texttt{VBMicroLensing}. Because these implementations employ fundamentally different hardware architectures (GPU versus CPU), we have reported normalized relative timings rather than direct wall-clock comparisons in Figure~\ref{fig:time_compare}. A more detailed discussion of the uniform-source hardware benchmarks is provided in \citet{Twinkle2025}.}
\revisionii{Finally, it is worth noting that for the specific \texttt{AdaptiveContouring} implementation, the true numerical error is evaluated \textit{a posteriori} to be consistently about two orders of magnitude smaller than the user-specified control parameter. While this operational characteristic shifts its performance curve horizontally, the scaling slope (the power-law exponent) experiences only minor changes. We retain the input tolerance as the horizontal axis across all methods to consistently demonstrate how the computational cost scales with the precision explicitly requested by the user.}

\response{For concentric-disk methods, the computational cost of achieving higher precision comprises two factors: the time required to compute the magnification of each uniform disk, and the total number of concentric disks needed. As the target error decreases, evaluating each uniform source takes longer, and the required number of integration nodes (disks) simultaneously increases. Because \texttt{Twinkle} and \texttt{VBMicroLensing} utilize similar underlying algorithms for uniform sources, their uniform-source scaling exponents are comparable ($\approx-0.2$). For limb-darkened sources, the total time is roughly the product of the single-disk time and the total number of disks. Therefore, the ``extra'' power-law exponent required for limb darkening—which reflects the convergence rate of the numerical integration—is $-0.22$ for \texttt{Twinkle}, $-0.42$ for \texttt{VBMicroLensing}. The reciprocals of these exponents ($\approx-4.5$ and $\approx-2.4$) correspond directly to the convergence rates of the limb-darkening integrals as a function of $N_{uni}$ discussed in Section~\ref{sec:converge_behavior}, validating our theoretical expectations. The resulting shallower slope of our method indicates improved computational efficiency for high-precision calculations relative to the concentric-ring approach.}

\response{For the \texttt{AdaptiveContouring} method, the single boundary integral approach does not require repetitive evaluations of concentric sources, and thus does not exhibit this ``extra'' scaling exponent. However, because the integrand in this formulation is highly complex, high-order numerical quadrature schemes \revisionii{have not yet been established}. Consequently, its computational time scales with an exponent of approximately $-1.0$ \revisionii{for the parameter choices considered here}. }

\section{Discussion}\label{sec:discussion}

\subsection{Error Items Strategy}

In \S~\ref{sec:error}, we introduced two empirical error items, $E_{c}$ in Equation~\eqref{eq:error_cross} and $E_{\mathrm{ini}}$ in Equation~\eqref{eq:error_initial}; this subsection examines their validity.

\begin{figure*}
\centering
\includegraphics[width=\textwidth]{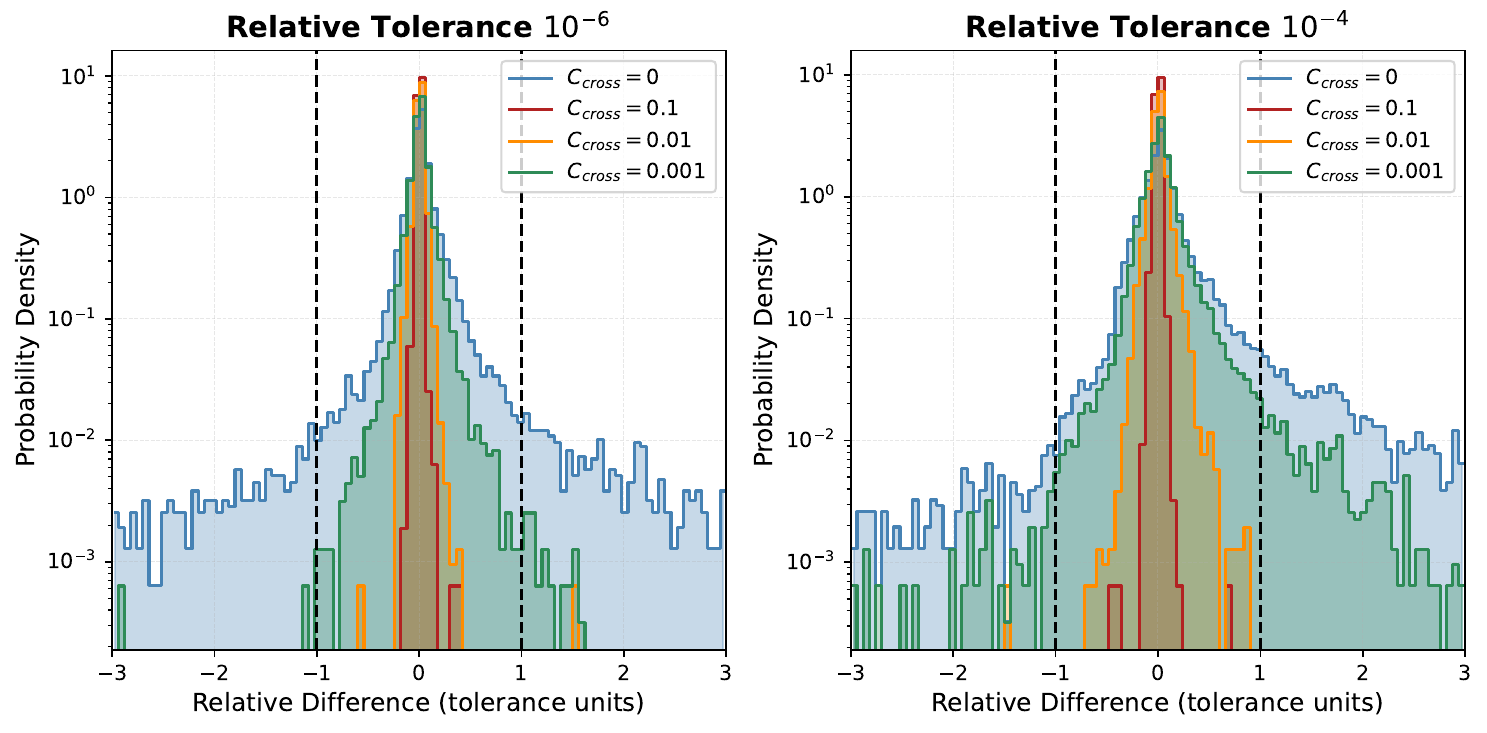}

\caption{Distribution of the relative magnification residuals for different choices of \revisionii{the coefficient} $C_{\mathrm{cross}}$ \revisionii{in Equation~\eqref{eq:error_cross}}. The left and right panels correspond to a relative tolerance of $10^{-6}$ and $10^{-4}$, respectively. The horizontal axis shows the residual normalized by the tolerance, and the vertical axis is the probability density. Histograms in different colors represent different values of $C_{\mathrm{cross}}$. The residuals are measured relative to the result obtained with $C_{\mathrm{cross}}=1$ (the conservative bound); $C_{\mathrm{cross}}=0$ corresponds to omitting the caustic-crossing error term entirely. The data comprise all caustic-crossing source positions from Figure~\ref{fig:2D_comparision}, totaling roughly $10^5$ points. As $C_{\mathrm{cross}}$ decreases, the dispersion of the residuals increases. The adopted value $C_{\mathrm{cross}}=0.1$ yields reliable results across both tolerance levels.}

\label{fig:C_cross}
\end{figure*}

For the caustic crossing error $E_{c}$, the error bound derived from the monotonicity Equation~\eqref{eq:ul_bounds} is a strict upper bound; hence $C_{\mathrm{cross}}=1$ provides a conservative, safe reference. To evaluate the practical choice of $C_{\mathrm{cross}}$ , we compare the relative magnification residuals obtained with different values of this coefficient. The test uses the same dataset as in Fig.~\ref{fig:2D_comparision}, which contains roughly $10^5$ source positions involving caustic crossings.

We consider two relative tolerance levels, $10^{-4}$ and $10 ^{-6}$. Figure~\ref{fig:C_cross} shows that omitting the caustic-crossing error term (or using a very small coefficient) leads to significant dispersion in the residuals, with a non-negligible fraction of points exceeding the prescribed tolerance. Both $C_{\mathrm{cross}}=0.01$ and $C_{\mathrm{cross}} = 0.1$ keep the vast majority of residuals within the tolerance. However, with $C_{\mathrm{cross}}=0.01$ about $10^{-5}$ of the points lie slightly outside the tolerance, whereas $C_{\mathrm{cross}}=0.1$ eliminates such outliers. The computational overhead introduced by the larger coefficient is only $10\%$ in caustic-crossing regions. Therefore, we adopt $C_{\mathrm{cross}} = 0.1$ as a robust compromise between accuracy and efficiency.
 
For the initial error, we compare two error forms $E_2$ and $E_1$. In most cases, $E_2$ is the dominant term, so \typo{$E_{\mathrm{ini}}$} naturally reduces to $E_2$ through the max operation in Equation~\eqref{eq:error_initial}. However, when the source crosses the caustic, near the cusp of the caustic, or inside the ``wings'' structure (visible in Figure~\ref{fig:2D_comparision_a}), $E_2$ could be much smaller and underestimate the initial error. The error item $E_c$ would protect the error estimation for the caustic crossing case, but the cusp case and wings case should be distinguished.

The “wings” correspond to positions where the magnification varies only weakly with radius; there, both $E_2$ and $E_1$ are small, and using $E_1$ correctly reflects the gentle behavior of the integrand. Near a cusp, however, $E_2$ can drop unexpectedly in particular locations while the actual error is governed by higher-order derivatives. In such locations $E_1$ remains large, providing a more conservative and reliable error bound.

The statistical behavior of the initial-step error estimator is examined using the dataset of Figure~\ref{fig:2D_comparision}. The ratio of the actual integration error to \typo{$E_{\mathrm{ini}}$} for \typo{$C_{\mathrm{ini}}=1$} has a mean of $0.024$ and a standard deviation of $0.003$. We therefore adopt \typo{$C_{\mathrm{ini}}=0.1$} to obtain a tighter, yet still conservative, error bound for the initial step. In applications that demand extremely high reliability, the safety can be further increased by imposing a minimum number of uniform-source evaluations $N_{\mathrm{uni}}$ before the integration is allowed to terminate.

\subsection{\response{Astrometry}}

\response{In addition to the magnification, limb darkening also affects the astrometric centroid shift. For a uniformly bright source of radius $R$, the centroid position $\Theta$ of the images can be calculated via the image area $\Omega$ and the first moment of area $X_0$:}
\begin{equation}
\begin{aligned}
    X_0 &= \int z \mathrm{d}\Omega 
    =  \int^R_0 \left(\int^{2\pi}_0  \sum_i \left[z(r,\theta)A_{\mathrm{point},i}(r,\theta) \right] \mathrm{d}\theta \right)  r \mathrm{d}r, \ \
        \Theta = \frac{X_0}{\Omega}. 
\end{aligned}
\end{equation}
\response{The integration region for the first moment $X_0$ is the interior of the images, with the complex number $z$ representing the position in the image plane as the integrand. The second equals sign transforms the integral from the image plane to the source plane. Since one single point $(r,\theta)$ in the source plane corresponds to multiple images $z$, a summation over each image point, weighted by point-image magnifications $A_{\mathrm{point},i}$, is required.}

\response{For a limb-darkened source, the centroid $\Theta_{\mathrm{LD}}$ depends on the limb-darkening profile $f$. Each position $z$ in the image plane can be mapped back to the source plane to obtain the corresponding limb-darkening factor $f\left[r(z)\right]$. Using this as the weight, we define the first moment for a limb-darkened source $X_{\mathrm{LD}}$. By analogy with Equations~\eqref{eq:M_Gamma} and \eqref{eq:area-integration}, it can similarly be expressed as an integral of the uniform-brightness first moment $X_0$:}
\begin{equation}
\begin{aligned}
    X_{\mathrm{LD}} &= \int z f\left[r(z)\right] \mathrm{d}\Omega \\
    &= \int^R_0 \left(\int^{2\pi}_0  \sum_i \left[z_i(r,\theta)A_{\mathrm{point},i}(r,\theta) \right] \mathrm{d}\theta \right)  r f(r) \mathrm{d}r \\
    &= \int^R_0 X_0'(r)f(r) \mathrm{d}r \\
    &= X_0(R)f(R) + \int^{f(0)}_{f(R)} X_0\left[r(f)\right] \mathrm{d} f,   \\
    \Theta_{\mathrm{LD}} &= \frac{\int z f(z) \mathrm{d}\Omega}{\int  f(z) \mathrm{d}\Omega} = \frac{X_{\mathrm{LD}}}{\pi R^2 A_{\mathrm{\Gamma}}}.
\end{aligned}
\end{equation}
\revisionii{In the second line, the subscript $i$ indexes the individual images produced by the microlensing effect. Because a single point source at $(r, \theta)$ generally maps to multiple images, each image has a specific position $z_i$ and a corresponding point-image magnification $A_{\mathrm{point},i}$. The term within the inner brackets therefore represents the magnification-weighted sum of the positions of these individual images.}
\response{In the third line, the first moment $X_0$ is expressed as a function of $r$, representing the first moment of a concentric source of radius $r$. The prime denotes differentiation with respect to $r$. Given that the magnification has already been computed, $X_0$ requires almost no additional computational cost \citep{VBBL3}. 
Compared to the boundary-integral method \citep[e.g.,][Sect. 4]{1998A&A...333L..79D}, which necessitates evaluating a different and complicated vector field for the first moment, this reuse of intermediate variables represents a significant computational advantage. 
Finally, the centroid position $\Theta_{\mathrm{LD}}$ of the limb-darkened source can be obtained from the limb-darkened first moment $X_{\mathrm{LD}}$ and the limb-darkened magnification $A_{\mathrm{\Gamma}}$.}

\subsection{Different Limb-darkening Profile}

This paper has focused on the widely used linear limb darkening profile, but the integral transformation given in Equation~\eqref{eq:area-integration} remains valid for any limb darkening profile. 
For example, the square root limb darkening profile $f_{\mathrm{SR}}$ as a function of the direction cosine $\mu$ parameterized by two factors $\Gamma$ and $\Lambda$ is also usually used \citep{An2002}:
\begin{equation}
\begin{aligned}
f_{\mathrm{SR}}(\mu) &=  1-\Gamma - \Lambda + \frac{3\Gamma}{2}\mu + \frac{5\Lambda}{4}\sqrt{\mu}.
\end{aligned}
\end{equation}
The integration argument $x=\sqrt{\mu}$ could perform better because it removes the singular behavior of the integrand near the limb, leading to more stable numerical quadrature. Therefore, Equation~\eqref{eq:area-integration} could be reformulated as:
\begin{equation}\label{eq:generalLD}
\begin{aligned}
f_{\mathrm{SR}}(x) &=  1-\Gamma - \Lambda + \frac{3\Gamma}{2}x^2 + \frac{5\Lambda}{4}x, \\
A_{\mathrm{SR}}(R) &= (1-\Gamma-\Lambda)A_0(R) +  \int^1_0 \tilde{{A}}_0(x^2)(1-x^4) \left(3\Gamma x + \frac{5\Lambda}{4}\right) \mathrm{d}x.
\end{aligned}
\end{equation}
The other profiles, including quadratic limb-darkening and logarithmic limb-darkening, can also be expressed in the integral form of Equation~\eqref{eq:area-integration}.

\section{Conclusion}

In this work, we have developed a new concentric-disk integration method for computing the magnification of linearly limb-darkened extended sources in gravitational microlensing. Through some mathematical transformations, the integration is reformulated into a form suitable for high-order numerical quadrature, leading to significantly accelerated convergence. This method has been successfully implemented within the \texttt{Twinkle} code. 
\response{In addition to photometric magnification, the same framework can be naturally extended to astrometric calculations.} 
Moreover, the method is general and can be readily applied to any microlensing code capable of providing the uniform-source magnification and the number of caustic crossings.

For linear limb darkening, the magnification integral is recast into the compact form of Equation~\eqref{eq:LD_Inte}. The magnification for an arbitrary linear limb-darkening coefficient $\Gamma$ can be expressed as a linear combination of the uniform-source magnification and the magnification for the specific case $\Gamma=1$. 
The computation of the latter term, $A_1$, is optimized by \revisionii{reformulating the limb-darkening integral through integration by parts and} adopting the direction cosine $\mu$ as the integration variable.
\revisionii{This reformulation enables the direct application of high-order adaptive quadrature schemes, providing a key advantage over concentric-ring formulations based on image area.}

Building on Equation~\eqref{eq:LD_Inte}, we employ an Adaptive Simpson's algorithm for numerical integration. To maintain robustness in regions where the magnification derivative changes abruptly, 
we introduce an additional error term (Equation~\eqref{eq:error_cross}) triggered by a change in the number of caustic crossings. This error estimate relies solely on the monotonicity of the integrand, ensuring that Simpson's rule can be applied safely even across intervals containing discontinuities in higher-order derivatives. 
\refinement{We further include a hidden-caustic error term (Equation~\eqref{eq:error_hidden}) to guard against undetected small caustic structures between adjacent annuli.}

We demonstrate the performance gain of the concentric-disk method 
\refinement{using representative binary-lens configurations.} 
In the vicinity of caustics, the concentric-disk method typically achieves the same accuracy using only about $30\%$ of the uniform-source magnification evaluations required by the concentric-ring approach. More fundamentally, the convergence rate improves from approximately $N^{-2}_{\rm{uni}}$ for the concentric-ring method to better than $N^{-4}_{\rm{uni}}$ for the concentric-disk method. This enhancement in the power-law exponent is particularly useful for high-precision computations, making the algorithm especially valuable for modeling high-quality photometric data.

Finally, we note that the linear limb-darkening integration framework presented here is not restricted to binary-lens systems; it is applicable to any lens configuration because it operates on the source brightness profile independently of the specific lens geometry. 
Moreover, the transformation leading to Equation~\eqref{eq:area-integration} remains valid for \refinement{more general} limb-darkening profiles. Although the resulting integral could become more complex \refinement{(e.g., Equation~\eqref{eq:generalLD})}, high-order integration schemes like Simpson's method can still be employed. This generality ensures that \texttt{Twinkle} limb-darkening integration method can be reliably applied to a wide range of complicated microlensing events.

\section*{Acknowledgements}

We thank the reviewer for constructive comments. We are grateful to Subo Dong for detailed discussions and valuable suggestions on the manuscript. We also thank Lile Wang for useful comments. This work is supported by the National Natural Science Foundation of China (Grant No. 12573067).

\bibliography{sample701}{}
\bibliographystyle{aasjournalv7}



\end{CJK*}
\end{document}